\documentclass[12pt,preprint]{aastex}

\usepackage{epsfig}
\usepackage{graphicx}
\usepackage{epstopdf}
\usepackage{amsmath}
\usepackage{amssymb}
\usepackage{appendix}

\def\be{\begin{equation}}
\def\ee{\end{equation}}
\def\ba{\begin{eqnarray}}
\def\ea{\end{eqnarray}}
\def\go{\mathrel{\raise.3ex\hbox{$>$}\mkern-14mu
             \lower0.6ex\hbox{$\sim$}}}
\def\lo{\mathrel{\raise.3ex\hbox{$<$}\mkern-14mu
             \lower0.6ex\hbox{$\sim$}}}

\begin{document}

\title{ Magnetar Giant Flares in Multipolar Magnetic Fields --- \\
I. Fully and Partially Open Eruptions of Flux Ropes }

\author{Lei Huang\altaffilmark{1,3,4} and Cong Yu\altaffilmark{2,4}}
\altaffiltext{1}{Key Laboratory for Research in Galaxies and
Cosmology, Shanghai Astronomical Observatory, Chinese Academy of
Sciences, Shanghai, 200030, China; {\tt muduri@shao.ac.cn}}
\altaffiltext{2}{Yunnan Observatories, Chinese Academy of
Sciences, Kunming, 650011, China; {\tt cyu@ynao.ac.cn}}
\altaffiltext{3}{Key Laboratory of Radio Astronomy, Chinese
Academy of Sciences, China. } \altaffiltext{4}{Key Laboratory for
the Structure and Evolution of Celestial Object, Chinese Academy
of Sciences, Kunming, 650011, China; }

\begin{abstract}
We propose a catastrophic eruption model for magnetar's enormous
energy release during giant flares, in which a toroidal and
helically twisted flux rope is embedded within a force-free
magnetosphere. The flux rope stays in stable equilibrium states
initially and evolves quasi-statically. Upon the loss of
equilibrium point is reached, the flux rope cannot sustain the
stable equilibrium states and erupts catastrophically. During the
process, the magnetic energy stored in the magnetosphere is
rapidly released as the result of destabilization of global
magnetic topology. The magnetospheric energy that could be
accumulated is of vital importance for the outbursts of magnetars.
We carefully establish the fully open fields and partially open
fields for various boundary conditions at the magnetar surface and
study the relevant energy thresholds. By investigating the
magnetic energy accumulated at the critical catastrophic point, we
find that it is possible to drive fully open eruptions for dipole
dominated background fields. Nevertheless, it is hard to generate
fully open magnetic eruptions for multipolar background fields.
Given the observational importance of the multipolar magnetic
fields in the vicinity of the magnetar surface, it would be
worthwhile to explore the possibility of the alternative eruption
approach in multipolar background fields. Fortunately, we find
that flux ropes may give rise to partially open eruptions in the
multipolar fields, which involve only partial opening up of
background fields. The energy release fractions are greater for cases with central-arcaded
multipoles than those with central-caved multipoles emerged in background fields.
Eruptions would fail only when the centrally-caved multipoles become extremely strong.
\end{abstract}



\section{Introduction}
Two small subsets of neutron stars $-$ anomalous X-ray pulsars
(AXPs) and soft gamma-ray repeaters (SGRs) are interpreted as
magnetars, which are neutron stars endowed with ultra-strong
($10^{14}-10^{15}$G) magnetic fields
\citep{Maze79,MS95,Kouv98,GKW02}. The magnetic energy dissipation
is commonly believed to account for their high energy persistent
emissions and spasmodic outbursts \citep{DT92,WT06,Mere08}. One of
the most intriguing phenomena related to magnetars is the episodic
giant flare, which involves tremendous magnetic energy release
\citep{Hurl05}. The physical origin of giant flares still remains
the bewildering enigma in high energy astrophysics. The
magnetospheric eruption models
\citep[e.g.,][]{Lyut06,GH10,Yu12,YH13} can naturally explain the
short timescale, $\sim0.25$ms, of the rise time in giant flares
\citep{Palm05}, although the precise mechanism of the eruption is
still under debate \citep{Mere13}. It is worthwhile to note that,
in the magnetospheric eruption model, the magnetosphere stays in a
stable equilibrium state at the pre-eruptive stage, in which the
magnetosphere evolves quasi-statically. As a result, the magnetic
energy released in an eruption is gradually accumulated on a
timescale much longer the dynamical timescale of giant flares.
Since the energy accumulation takes place very gradually, the
question how such long timescale events could initiate the sudden
energy release, i.e., giant flares, on a very short dynamical
timescale, remains elusive for the magnetospheric eruption model.

To resolve this puzzle, a catastrophic flux rope eruption model
has been put forward to explain the magnetar giant flare
\citep{Yu12,YH13}. In this particular scenario, the giant flare is
considered to be driven by the destabilization of large-scale
magnetospheric magnetic fields rather than the abrupt fracture of
the neutron star crust. The most distinguishing feature in our
model is that a helically twisted flux rope is embedded within the
magnetosphere. Magnetic flux ropes could be naturally generated
due to the the magnetic helicity injections from the magnetar
interior (Thompson et al .2002; Lyutikov 2006; G\"{o}tz et al.
2007; Gill \& Heyl 2010).
Such flux ropes are also an indispensable ingredient to explain
the radio afterglow of SGR1806 \citep{Gaen05,Lyut06}. It is
interesting to note that the interior of the flux rope is
helically twisted. The magnetic twist is locally confined within
the flux rope, which is at variance with the global twist proposed
by recent authors \citep{Lyut13,Parf13}. These locally twisted
magnetic features, when compared to the global twist, seem to be
more relevant to the recent observations
\citep[e.g.,][]{Wood07,PG08}.

The flux rope eruption model tries to resolve a primary issue
concerning the trigger mechanism of the magnetar eruption. The
most appealing characteristic is that, during the flux rope's
evolution, it could make the catastrophic state transition from
stable equilibrium states to unstable equilibrium states
spontaneously in accordance with the variations at the neutron
star surface. The emergence of flux from the interior \citep{KR98,GMH07}
and/or the shuffling of the crust \citep{Rude91} causes the flux rope to
evolve to a critical loss of equilibrium point, beyond which no
stable equilibrium state can be sustained and the flux rope erupts
abruptly.  It is widely accepted that the energy is progressively
accumulated in an initially closed force-free field before the
flux rope reaches the critical loss of equilibrium point. The flux
rope's subsequent catastrophic eruption, beyond this particular
point, leads to the opening up of initially closed magnetic field
configuration as well as huge energy release. Since the stored
energy prior to an eruption is determined by the intrinsic
properties of the magnetosphere rather than the tensile strength
of the crust \citep{Yu12}, a fundamental question is raised
naturally as to whether the magnetic energy in the force free
magnetosphere could build up enough energy to support eruptive
giant flares.

The fierce eruptive event is thought to open up the pre-eruptive,
originally closed magnetic field lines. Observationally, in the
post-eruption epoch of giant flares, magnetic field configurations
are indeed inferred to stretch outwards to form an open field
structure \citep[e.g.][]{Wood01}. From the perspective of energy
budget, a closed field configuration is physically favored for
giant flares only if the magnetic energy stored in magnetosphere
can exceeds that stored in an open field configuration. In other
words, in addition to power the giant flares, the magnetic free
energy has to be able to open up the initially closed magnetic
fields. This constitutes a serious bottleneck for magnetar giant
flares because it is realized that to open up the initially closed
field lines requires considerable extra work to be done
\citep{Aly84,Aly91,Stur91,Yu12}. To get over the energy threshold
constrained by the post-eruptive open field configurations, the
pre-eruptive magnetosphere must accumulate more energy in excess
of the threshold set by the open field configurations.
It should be pointed out that the field lines can also be opened up
at a greater distance from the star by strong neutron star winds
\citep[e.g.,][]{Bucc06}. How the wind affect the magnetar eruption
energetics is still an open issue. For simplicity, we only
consider the case in which the field lines are opened up by the
process of magnetic flux injection from the neutron star interior.

The observational feature of strong four-peaked pattern in the
pulse profile of the 1998 August 27 event from SGR 1900+14
indicates that the geometry of the magnetic field was quite
complicated in regions close to the star \citep{Fero01}.
Recent calculations also show that multipolar magnetic fields may
also have important effects on the emission of the magnetars \citep{Pava09}.
As a result, it is reasonable to infer that,
in the very vicinity of the magnetar surface, the field
configuration involves higher multipoles. The electric currents
formed during the birth of magnetars slowly push out from within
the magnetar and generate active regions on the magnetar surface.
These active regions manifest themselves as the multipolar regions
on the magnetar surface.
The flux rope eruption in multipolar background fields, unlike the
behavior in dipole background fields, may just involve opening-up
of part of the closed magnetic flux systems. Thus, it is
interesting to explore the physical behavior of the flux rope in
response to these more complex boundary conditions. However, for
more complex boundary conditions, no solid investigations about
the energy threshold specified by the fully/partially open field
configurations have been performed. A related question, which is
of crucial significance for the physical feasibility of the flux
rope eruption model, i.e., whether the flux rope could build up
enough energy to drive the giant flare with these complex boundary
conditions, also remains to be answered.

In this paper, we will establish a force-free magnetosphere model
with a helically twisted flux rope and examine the physical response
of the flux rope to the variations of the background multi-polar magnetic fields.
We perform rigorous calculations on the magnetic energy
accumulation in magnetospheres in various configurations.
Specifically, boundary conditions containing the dipolar term and
the high order multipolar term are considered in the background
closed magnetic field. Two kinds of open configurations, the
partially open and fully open fields, are considered to provide
energy thresholds for eruptions.
This paper is structured as follows. The model of pre-eruptive
magnetospheres with flux ropes and the multipolar boundary
conditions are introduced in Section 2. The physical behavior of
the pre-eruptive flux rope is described in Section 3, including
the equilibrium constraints and catastrophic loss of equilibrium
of the flux rope. The energetics of the flux rope in the
multipolar background fields are investigated in Section 4.
Conclusions and discussions are provided in Section 5.


\section{Pre-eruptive Force-Free Magnetospheres With Embedded Flux Ropes}

\subsection{Force-Free Magnetospheres with Helically Twisted Flux Ropes}

In our model, the most distinctive characteristic is that there
exists a toroidal and helically twisted flux rope in the
pre-eruptive magnetosphere. It is possible that the precursor
activity of a giant flare may inject certain amount of magnetic
helicity into the magnetosphere and generate such toroidal and
helically twisted flux ropes \citep{TLK02,GMH07,GH10}.
The toroidal flux rope has a major radius, $h$, which can be also
understood as the height of the flux rope measured from the
magnetar center, and a minor radius, $r_0$, which is small
compared to $h$. The magnetic twist of the flux rope is confined
within the flux rope, which is unlike the globally twisted
magnetic fields configurations proposed by some authors \citep{TLK02,Belo09,Parf13,Lyut13}.
These authors considered a non-potential force-free field
where the electric currents flow through the entire magnetosphere.
In comparison, our model only contains an electric current in a
spatially restricted region, viz, inside the helically twisted
flux rope. The magnetic field generated by the current inside the
flux rope can be represented by a wire carrying a current $I$ at
the center of the flux rope \citep{FP95}.

\begin{figure}
\includegraphics[scale=0.9]{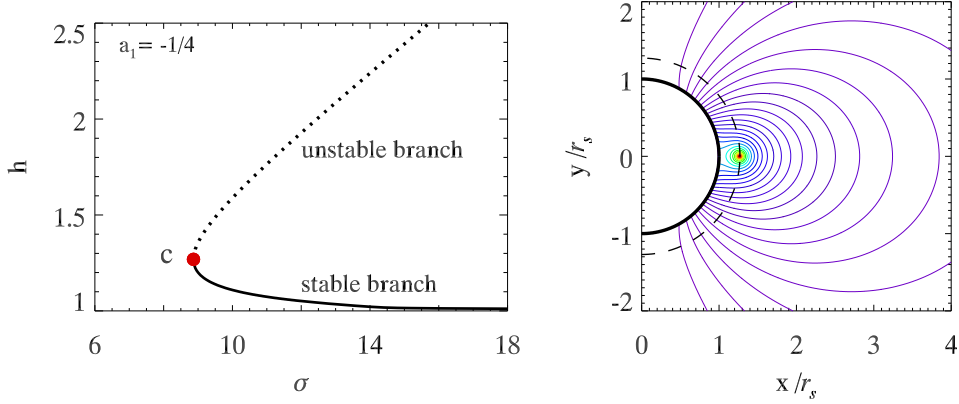}
\caption{\label{M0} {\it Left}: Equilibrium curve of the flux rope
in a dipole-dominated background with $a_1=-1/4$ and
$r_{00}=0.01$, which shows the dependence of the height of the
flux rope on the magnetic flux at the magnetar surface. The
magnetospheric field configuration . The lower stable branch and
the upper unstable branch are shown in solid line and dotted line,
respectively. The red dot marks the critical point.
{\it Right}: Magnetosphere with an embedded flux rope in the
pre-eruptive critical state. The thick solid semi-circle
represents the magnetar surface. The dashed line represents a
circle with a radius of the critical height of the flux rope $h_c$.
 }
\end{figure}

The presence of the flux rope separates the magnetosphere into two
regions, one is the region inside the flux rope. Further details
about the solution inside the flux rope are discussed in \citet{Yu12}.
The other is the region outside the flux rope, in which
the steady state axisymmetric magnetic field $\mathbf{B}$ takes
the following form in spherical coordinates $(r,\theta,\phi)$
\begin{eqnarray}\label{BrBtheta}
\mathbf{B} &=& -\frac{1}{r^2}\frac{\partial\Psi}{\partial\mu}\
\hat{\mathbf{e}}_r \ -\frac{1}{r\sin\theta}
\frac{\partial\Psi}{\partial r}\ \hat{\mathbf{e}}_\theta \ ,
\end{eqnarray}
where $\Psi(r,\mu)$ is the magnetic stream function and
$\mu=\cos\theta$. Here $\hat{\mathbf{e}}_r$ and
$\hat{\mathbf{e}}_\theta$ are the unit vector along the radial and
latitudinal direction, respectively. The force-free condition can
be expressed in terms of the Grad-Shafranov (GS) equation, which
explicitly reads ,
\begin{eqnarray}
\label{GS} \frac{\partial^2\Psi}{\partial r^2} +
\frac{(1-\mu^2)}{r^2}\ \frac{\partial^2\Psi}{\partial\mu^2}\ =\
-r\sin\theta\ \frac{4\pi}{c}J_\phi\ ,
\end{eqnarray}
where $c$ is the speed of light and the current density $J_\phi$
on the right hand side in this inhomogeneous GS equation is
induced by the toroidal flux rope, which is of the following form \citep{PF00,Yu12}
\begin{equation}
    J_\phi = \frac{I}{h}\ \delta(\mu) \ \delta(r-h)\ ,
\end{equation}
where $I$ designates the electric current in the flux rope.
According to the variable separation method, the general solution
to the GS equation can be conveniently written as
\begin{eqnarray}\label{general}
\Psi(r,\mu) &=& \sum_{i=0}^\infty
\bigg[c_{2i+1}R_{2i+1}(r)+d_{2i+1}r^{-2i-1}\bigg]\
\left[\frac{P_{2i}(\mu)-P_{2i+2}(\mu)}{4i+3}\right]\ ,
\end{eqnarray}
where $P_{2i}(\mu)$ is the Legendre polynomial and $R_{2i+1}(r)$
is a continuous function of $r$ \citep[see][]{Yu11b,Yu12}. The
coefficients $c_{2i+1}$'s are determined by the current inside the
flux rope. Once the magnetar surface boundary conditions are
fixed, the coefficients $d_{2i+1}$'s can be readily specified in
terms of $c_{2i+1}$'s and the boundary conditions. More technical
details to obtain solutions of the GS equation can be found in \citet{Yu12}.
Once we obtain the spatial distribution of the magnetic
stream function, the magnetic field configuration in the
magnetosphere can be determined. Illustrative figures of the
magnetic field configurations are shown in the panel $b$ of Fig.\ref{M0} and Fig.\ref{M1d5}.
The height of the flux rope (shown as dashed line)
in these two fiugres is 1.27$r_s$ and 2.20$r_s$, respectively
($r_s$ is the magnetar radius, shown in thick solid line in these
figures). In this paper we find that boundary conditions have
important influences on the flux rope eruptions and in the
following section we will further discuss the boundary conditions
we adopt.

\subsection{Multipolar Boundary Conditions and Post-Eruptive Energy Thresholds}

The GS equation is solved in the range $[r_s, \infty)$, where
$r_s$ is the magnetar radius. The boundary conditions both at
$r=r_s$ and $r\rightarrow\infty$ must be explicitly specified
before we solve the GS equation (\ref{GS}). The physical
requirement that $|\nabla\Psi| \rightarrow 0$ for $r\rightarrow
\infty$ can be trivially satisfied \citep{Yu12}. At the magnetar
surface $r=r_s$, we adopt the multipolar boundary condition as \citep[e.g.,][]{Anti99,ZL01}
\begin{equation}\label{Psi0}
\Psi(r_s, \mu) = \Psi_0 \sigma \Theta(\mu) \ ,
\end{equation}
where $\Psi_0$ is a constant with magnetic flux dimension and the
dimensionless variable $\sigma$ indicates the magnitude of
magnetic flux at the magnetar surface. The large scale field
configuration of the neutron star is essentially a dipole field.
However, in the vicinity of the neutron star surface, which is
exactly the location where the catastrophic loss of equilibrium
takes place, the magnetic field may be more complicated than a
simple dipole \citep{Fero01}. To model multipolar regions on
the neutron star surface, we include high order multipolar
components in addition to the dipole field and the angular
dependence of the function $\Theta (\mu)$ can be written
explicitly as
\begin{equation}
\label{pb}
    \Theta(\mu) = (1-\mu^2) +  a_1\ (5\mu^2-1)(1-\mu^2)\ .
\end{equation}
The first term $(1-\mu^2)$ denotes the dipolar component of the
magnetic fields and the additional term represents the
contributions from high order multipolar components. The parameter
$a_1$ determines the strength of the multipoles. The value of
$a_1$ can be either positive or negative, the schematic
illustration of the field configurations for different sign of
$a_1$ are shown in middle- and bottom-left panels of Fig.\ref{bound} in Appendix A.
In this paper, we confine $a_1$ in the range of $[-1,1]$. Note that larger
values of $|a_1|$ may indicate stronger magnetic activity of the
magnetar \citep{KR98,GMH07,Pava09}.

In the post-eruption epoch of giant flares, magnetic field lines
are stretched outwards to form open field structures \citep{Wood01}.
The possible post-eruptive open field configurations are
specified by the profile of flux distribution at the magnetar
surface. When the absolute value of the parameter $a_1$ is small,
the background magnetic field is basically dipolar. The
demarcation between dipole dominated fields and multipole
dominated fields is determined by how many extremum points exist
in the profile of the boundary flux distribution. More explicitly,
if the parameter $a_1$ is in the range of $[-1/4,1/6]$, only one
extremum point exits at $\mu=0$ in the boundary flux profile and
the background magnetospheric field is essentially dipole
dominated.
Otherwise, the boundary flux function we take has three extremum
points at $\mu_0=0$, $\mu_1=\sqrt{(6a_1-1)/10a_1}$, and
$\mu_2=-\mu_1$, respectively (see Fig.\ref{bound} in Appendix A).
Due to the existence of the multiple extremum points, the
multipolar configurations naturally arise in the background field.
If $a_1\in(1/6,1]$, the background field has a central-caved
profile on the boundary. If $a_1\in[-1,-1/4)$, the
background field has a central-arcaded profile on the
boundary. Note that for a dipole dominated background, there is
only a single closed flux system in the magnetosphere
(see in top-left panel in Fig.\ref{bound}.).
The fully opening up of magnetic field specifically means the
opening up of this particular closed flux system. However for a
multipole dominated background, there are multiple closed flux
systems in the magnetosphere (see in middle- and bottom-left panels in Fig.\ref{bound}.).
The fully opening up of magnetic fields surely involves
all the closed flux systems. However, it is possible that only
part of the closed flux systems opens up and the post-eruptive
field is then called partially open field. Details to obtain both
the fully and partially open field configurations according to the
boundary conditions we adopt are further discussed in Appendix A.

Throughout this work, the magnetic energy is scaled by the energy
of the potential field satisfying $\nabla\times\mathbf{B}=0$,
which has the minimum magnetic energy and is denoted by $W_{\rm
pot}$. The magnetic energy of fully and partially open fields are
denoted by $W^{\rm f}_{\rm open}$ and $W^{\rm p}_{\rm open}$,
respectively. Further descriptions on how to calculate the
magnetic energy of these two open states are given in Appendix B.
The fully and partially open fields constitutes two energy
thresholds, $W^{\rm f}_{\rm open}/W_{\rm pot}$ and $W^{\rm p}_{\rm
open}/W_{\rm pot}$, for the flux rope eruptions.
It is conceivable that the flux rope must accumulate enough
magnetic energy to either fully or partially open up the initially
closed fields.
More specifically, the magnetic energy stored in the critical
pre-eruptive state, $W_{\rm pre}(h_c)/W_{\rm pot}$, should be
larger than the fully or partially open threshold. Note that the
magnetic energy of the critical pre-eruptive state, $W_{\rm
pre}(h_c)$, is closely related to the catastrophic behavior of the
flux rope, which will be discussed in the following section.

\section{Catastrophic Response of Flux Ropes to Variations at Magnetar Surface}

It should be pointed out that, at the pre-eruptive stage, the flux
rope stays in a stable equilibrium state and evolves
quasi-statically with variations at the neutron star surface.
During the pre-eruptive stage, the flux rope is unable to erupt
and the magnetic energy is gradually accumulated in the magnetosphere.
Upon the the critical loss of equilibrium point is reached, the
quasi-static evolution is replaced by the subsequent dynamical
evolution \citep{Yu12,YH13}. The accumulated energy at the
catastrophic loss of equilibrium point is of particular
significance. This is because, beyond this point, no further
gradual energy buildup is allowed and the flux rope's dynamic
behavior should be supported by the accumulated energy at this
point. We denote the  pre-eruptive energy at this loss of
equilibrium point as $W_{\rm pre}$. To calculate the pre-eruptive
state energy, $W_{\rm pre}$, it is necessary to know when the flux
rope begins to lose its equilibrium.

\subsection{Equilibrium Constranits of Flux Ropes}

We adopt the the \citet{Lund50} force-free solution to represent
the current density and field inside the toroidal flux rope. This
solution, though originally derived for straight cylindrical
twisted flux rope, is still valid as long as the minor radius
$r_0$ is much smaller than the major radius, $h$. The axial
magnetic flux conservation of the flux rope suggests that the
minor radius $r_0$ is inversely proportional to the current
carried by the flux rope $I$ (Yu 2012)
\begin{equation}
r_0 = r_{00} I_0 / I = r_{00}/J \ ,
\end{equation}
where the dimensionless current $J$ is defined by $J\equiv I/I_0$.
The scaling current $I_0 =\Psi_0 c/r_s$ is determined by the
magnetic flux constant $\Psi_0$ in Equation (\ref{Psi0}), the
magnetar radius $r_s$ and the speed of light $c$. For numerical
conveniences, we scale the length by the magnetar radius $r_s$,
magnetic flux by $\Psi_0$ and current by $I_0$ in our following
calculations. The parameter $r_{00}$ is the value of $r_0$ when
$J=1$. Typically for a flux rope with minor radius of $0.1$km, the
value of the parameter $r_{00}\sim 0.01$ (We adopt the typical
neutron star radius $r_s\sim10$km).

In what follows, we consider slow responses of the flux rope to
changes at the magnetar surface, and thus the flux rope is assumed
to stay in a quasi-static equilibrium state on a sufficiently long
timescale. Two aspects of the equilibrium constraint are
considered, i.e., the force balance condition and the ideal frozen-flux condition.
The force balance condition is satisfied when the total force
exerted on the flux rope vanishes. The current inside the flux
rope provides an outward force. The magnitude of this force is
equal to the current, $I$, times the magnetic field, $B_s$ \citep{Shaf66}:
\begin{equation}
B_s = \frac{I}{c h}\left( \ln \frac{8h}{r_0} - 1\right) \ ,
\end{equation}
where $h$ and $r_0$ are the major and minor radius of the flux
rope, respectively. This current-induced force must be balanced by
the external magnetic field $B_e$. To calculate the external
magnetic field $B_e$, the contribution from the current inside the
flux rope must be excluded \citep{Yu12}. Finally we can arrive at the
mechanical force balance condition
\begin{equation}\label{mechanical}
f(\sigma, J, h) \equiv B_s - B_e = 0  \ ,
\end{equation}
where the function $f(\sigma, J, h)$ can be written explicitly as
\begin{equation}\label{fdef}
f(\sigma, J, h) = \frac{J}{h}\left(\ln\frac{8Jh}{r_{00}}-1\right)
- \sum_{i=0}^\infty (2i+1) \left[
\frac{P_{2i}(0)-P_{2i+2}(0)}{4i+3}\right]
\frac{d_{2i+1}}{h^{2i+3}} \ . \nonumber
\end{equation}
We use $\sigma$ to represent the dimensionless magnitude of magnetic
flux at the magnetar surface, $J$, the dimensionless current and
$h$, the dimensionless height of the flux rope. Here the
coefficients $d_{2i+1}$'s are already determined in Equation
(\ref{general}). Note that $P_{2i}(0)$ and $P_{2i+2}(0)$ are
values of the Legendre polynomials at $\mu=0$.

The ideal frozen-flux condition must be also satisfied for the
magnetic stream function, $\Psi$. It demands the stream function
on the edge of the flux rope keep constant during the system's
evolution, which provides a connection between the variations at
the magnetar surface and the current flowing inside the flux rope.
Explicitly this condition can be written in the following form
\begin{equation}\label{frozen}
g(\sigma,J,h) \equiv \Psi(h-r_0,0) = {\rm const} \ ,
\end{equation}
where the function $g(\sigma, J, h)$ is defined by
\begin{equation}\label{gdef}
g(\sigma,J,h) = \sum_{i=0}^\infty
\left[\frac{P_{2i}(0)-P_{2i+2}(0)}{4i+3}\right] \left[ c_{2i+1}
\left(1-\frac{r_{00}}{Jh}\right)^{2i+2} + d_{2i+1}
\left(h-\frac{r_{00}}{J}\right)^{-2i-1} \right]\ . \nonumber
\end{equation}
Note that the symbols in this equation has the same meaning as
those in Equation (\ref{fdef}) and the coefficients $c_{2i+1}$'s
and $d_{2i+1}$'s are specified in Equation (\ref{general}). With
these two equations, the current $J$ and the height of the flux
rope $h$ can be determined numerically according to the
Newton-Raphson method for any given value of $\sigma$. In the
following we will investigate how the equilibrium height of the
flux rope behaves with the variations at the magnetar surface.

\subsection{Loss of Equilibrium of Flux Ropes in Response to Surface Variations}

The loss of equilibrium of the flux rope is triggered by slow
changes at the magnetar surface. There are two possible long
timescale processes that could occur at the magnetar surface. One
is that new magnetic flux, driven by plastic deformation of the
neutron star crust, may be injected continuously into the magnetosphere \citep{KR98,TLK02,LL12}.
Another interesting possibility is the crust horizontal movement \citep{Rude91,Jone03}.
The second possibility has been investigated in \citet{Yu12}. Here we only
consider the effects of flux injection on the behavior of the flux
rope for simplicity. As new magnetic fluxes are injected, the
background magnetic field would vary gradually. The background
magnetic field would decrease (increase) if the opposite (same)
polarity flux is injected.  Note that there exist two possible
field configurations, inverse and normal \citep{Yu12}. In the normal
configuration, the critical equilibrium height is rather low,
usually a few percent above the neutron star surface. Given the
regular arrangements occurring at the magnetar surface, the small
height of the normal configuration suggests that it may not
survive those arrangements at the magnetar surface \citep{Yu12}.
Hence we will focus on the inverse configurations in this paper.

To be specific we fix the value of $r_{00} = 0.01$ in the section.
The effects of varying $r_{00}$ will be further discussed in
section 4. By solving Equations (\ref{mechanical}) and
(\ref{frozen}), we can get the flux rope's equilibrium curves,
which show the variations of the flux rope's equilibrium height in
response to the gradual background magnetic flux changes. The
relevant results are shown in the panel a in Fig. 1 and Fig. 2. We
show a dipole dominated background field with $a_1=-1/4$ in Fig. 1
and a multipole dominated field with $a_1=2/3$ in Fig. 2. It can
be found that each equilibrium curve contains two branches, the
lower stable branch (solid line) and upper unstable branch (dotted
line). On the lower equilibrium branch, the total force on the
flux rope $F\propto I h (B_s-B_e)$ shows a negative derivative
with respect to $h$, i.e., $dF/dh<0$ \citep{Forb10}. Physically
speaking, the flux rope is stable if it lies on this branch, since
a slight upward displacement would create an inward restoring
force. The upper equilibrium branch, on the contrary, is unstable.
Since the total force shows a positive derivative, i.e.,
$dF/dh>0$, and a slight upward displacement on the flux rope lying
on this branch would generate an outward driving force. The two
branches are joined together at the critical loss of equilibrium
point (shown as red dot in these figures). In the left panel of
Fig. 1 and Fig. 2, we find that, with the decrease of the
parameter $\sigma$, the flux rope gradually approaches this
critical point. At this point, the flux rope can not sustain the
stable equilibrium any longer and erupt catastrophically. The
quasi-static evolution of the flux rope is then replaced by the
subsequent dynamical evolution.

In the right panel of the two figures, we show the
critical pre-eruptive magnetic field configuration, which corresponds to
the state represented by the red dot in the left panel. The
magnetar surface is shown in thick solid semi-circle and the
critical height of the flux rope is shown in a dashed circles with
a radius $r=h_c$ in each case. The critical heights for Fig. 1 and
Fig. 2 are about $1.27r_s$ and $2.20r_s$, respectively. The
possible post-eruptive configurations for the critical
pre-eruptive states in Fig. 1 and Fig. 2 are shown in Appendix A.
Note that whether the transition from the closed pre-eruptive
state to the fully (or partially) open post-eruptive state is
feasible or not is determined by the energy relations between the
two states. When the magnetic energy accumulated at the critical
pre-eruptive state is higher than the relevant post-eruptive
state, such state transitions are physically favored. To check the
feasibility of fully or partially open eruptions, we need to know
the magnetic energy accumulated at the critical loss of
equilibrium point. The energetics of the flux rope, or the
feasibility of the state transition will be further discussed in
Section 4.

\section{Energetics of Fully and Partially-Open Flux Rope Eruptions}

We have already known that the flux rope presents a catastrophic
behavior from the previous section. This is consistent with the
observational characteristics of giant flares. The flux rope
initially stays on the stable branch and loses its equilibrium
after evolving to the critical loss of equilibrium point. The flux
rope eruption would be physically favored if it contains magnetic
energy over the fully or partially open field energy threshold,
$W^{\rm f}_{\rm open}$ or $W^{\rm p}_{\rm open}$. In the following
we will study the energetics of the flux rope eruption model and
answer the question whether the state transition  is possible or not.

\subsection{Magnetic Energy Accumulation Prior to Catastrophe}

To check whether the flux rope is able to drive giant flares or
not, it is crucial to know the total magnetic energy that it could
accumulates before the catastrophe. The total magnetic energy
accumulated in the pre-eruptive state is the sum of the free
energy and the potential magnetic energy, i.e.,
$W_\mathrm{pre}=W_\mathrm{free}+W_\mathrm{pot}$. The free magnetic
energy of the system prior to catastrophe is equal to the work
required to move the flux rope from infinity to the location where
the flux rope lies. Thus the free magnetic energy is given by
\begin{eqnarray}
W_\mathrm{free} &=& -\int^h_\infty F dh^{\prime} = -\int^h_\infty
2\pi \ \frac{I h^{\prime}}{c} \ (B_s-B_e)\ dh^{\prime} \ ,
\end{eqnarray}
where $F$ is the total force exerted on the flux rope. The
potential magnetic energy $W_\mathrm{pot}$ of the magnetar is
\begin{equation}
\label{pot} W_\mathrm{pot} = \int \frac{{\bf B}_{\rm
pot}^2}{8\pi}\ dV = \int _{\partial V} \frac{B_{\rm pot}^2}{8\pi}
({\bf r} \cdot d{\bf S}) - \frac{1}{4\pi} \int_{\partial V} ({\bf
B}_{\rm pot}\cdot {\bf r}) ({\bf B}_{\rm pot}\cdot d{\bf S})\ ,
\end{equation}
where the volume integral is performed over the entire
magnetosphere outside the magnetar, $\bf r$ is the position vector
and $d{\bf S}$ is the surface area element directed outwards. Note
that in Equation (\ref{pot}) the volume integral has been already
transformed to the surface integral according to the magnetic
virial theorem \citep{Chandra61}.
In the above equation, the potential magnetic field ${\bf B}_{\rm
pot}$ can be obtained from the potential stream function
$\Psi_{\rm pot}$ via Equation (\ref{BrBtheta}). According to the
boundary condition, i.e., Equation (\ref{Psi0}), the potential
stream function $\Psi_{\rm pot}$ can be explicitly written as
\begin{eqnarray}
\Psi_{\rm pot}(r,\mu) &=& \Psi_0\sigma\ \left[ \frac{r_s}{r} +
a_1\ (5\mu^2-1)\frac{r_s^3}{r^3} \right]\ (1-\mu^2)\ .
\end{eqnarray}
The total magnetic energy accumulated in a pre-eruptive state is
the sum of the free energy and the potential magnetic energy,
i.e., $W_\mathrm{pre}=W_\mathrm{free}+W_\mathrm{pot}$.
By examining the pre-eruptive energy accumulation process, we are
able to figure out whether certain types of pre-eruptive states
may support the giant flare or not. In the following we will
investigate the energy accumulation process of the flux rope in
greater details.

\subsection{Fully Open Eruptions in Dipole Dominated Backgrounds}
The energy accumulation processes of the flux rope in the dipole
dominated and multipole dominated background are shown in the left
and right panel of Fig. 3, respectively. These two panels
correspond to the energy accumulation processes before the
catastrophic loss of equilibrium shown in Fig. 1 and Fig. 2,
respectively. In this section we focus on the energy accumulation
process in a dipole dominated background field shown in the left
panel. Since there is only one closed flux system in the dipole
dominated background for the case with $a_1=-1/4$, the opening up
of closed background always induces fully open eruptions. Fully
open magnetic fields can be obtained by flipping the boundary flux
distributions in the southern hemisphere, i.e., $\mu\in[-1,0]$. A
brief description of the procedures to obtain the field
configuration and the energy threshold constrained by the fully
open field is discussed in Appnedix A and B, respectively. The
fully open energy threshold for the case of $a_1=-1/4$ is shown in
dashed line in Fig. \ref{W0}. We also present in this figure the
normalized accumulated magnetic energy,
$W_\mathrm{pre}/W_\mathrm{pot}$, versus the background flux,
$\sigma$, before the catastrophe of flux ropes (in solid line).
Note that the energy reaches maximum at the critical pre-eruptive
point, marked by the red dot. Our calculations shows that the
accumulated energy of the critical pre-eruptive state, $W_{\rm
pre}(h_c)/W_{\rm pot}$\footnote{The parameter $h_c$ stands for the
critical height of the flux rope, the height where the red dot
lies.}, is $2.17$. The energy threshold constrained by the fully
open state, $W^{\rm f}_{\rm open}/W_{\rm pot}$, is $2.011$. The
energy release fraction of the pre-eruptive magnetosphere, $
[W_{\rm pre}(h_c) - W^{\rm f}_{\rm open}]/W_{\rm pre}(h_c)$, is
about $7\%$. Observationally, the total magnetic energy in the
magnetosphere is about $\sim 10^{46} ({\rm B}/10^{14}G)^2
(r_s/10{\rm km})^3$ergs, the giant flare is typically
$10^{44}$ergs, so only $1\%$ of magnetic energy release in the
magnetosphere could account for the giant flares
\citep{WT06,Mere08}. As a result, it is possible for the flux rope
to fully open up the background field and provides abundant
magnetic energy to drive a magnetar giant flare.

\subsection{Partially Open Eruptions in Multipolar Backgrounds}
\begin{figure}
\includegraphics[scale=0.9]{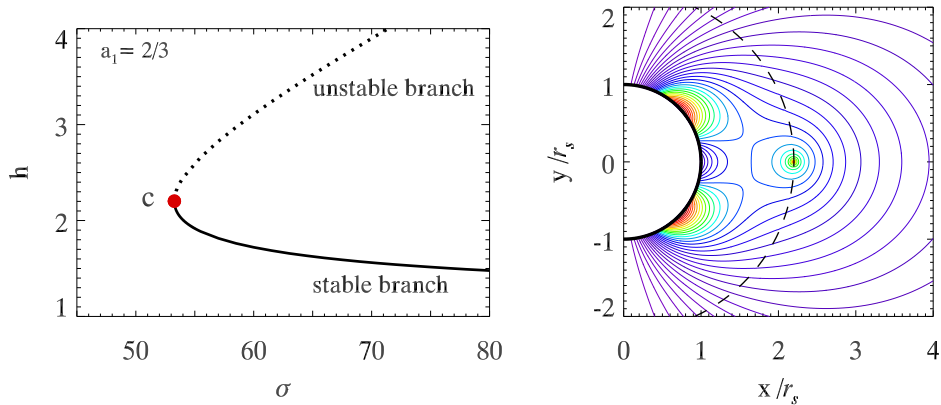}
\caption{\label{M1d5} The same as Fig.\ref{M0} but with a
multipole-dominated  background with $a_1=2/3$.
 }
\end{figure}

\begin{figure}
\includegraphics[scale=0.9]{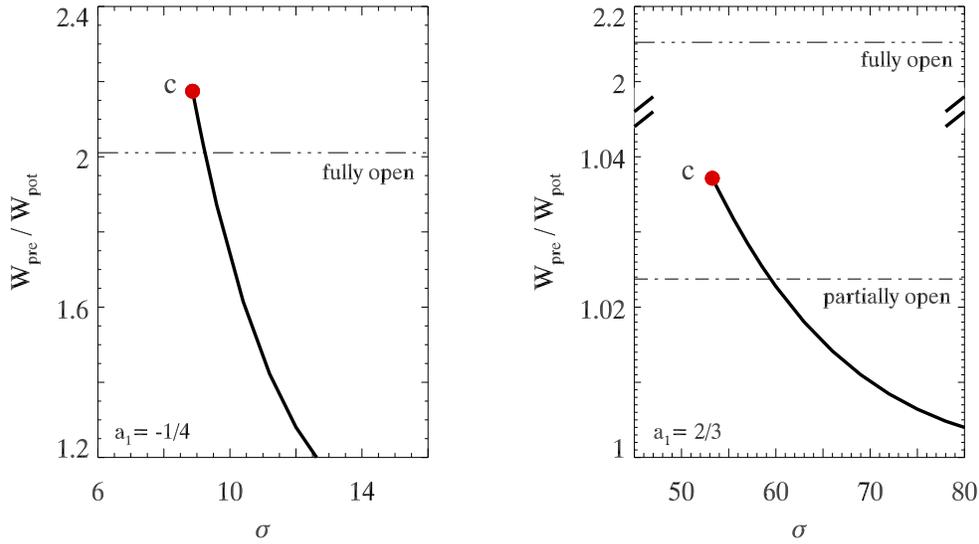}
\caption{\label{W0} {\it Left}: The energy accumulation process
before the catastrophe, $W_\mathrm{pre}/W_\mathrm{pot}$, in a
background field with $a_1=-1/4$, along the stable branch as shown
in Fig. \ref{M0}.
The fully open threshold is shown in dashed-three-dotted line. The
accumulated magnetic energy goes beyond the fully open field
threshold. {\it Right}: The energy accumulation before the
catastrophe in a background field with $a_1=2/3$, along the stable
branch as shown in Fig. \ref{M1d5}. The fully open and partially
open thresholds are shown in dashed-three-dotted and dashed-dotted
lines, respectively. The accumulated magnetic energy goes beyond
the partially open field threshold, but still lower than the the
fully open field threshold.
 }
\end{figure}

In the previous section, we have established that for a dipole
dominated background it is possible to induce fully open
eruptions. However, observation shows that multipolar magnetic
fields may be involved for magnetar giant flares
\citep{Fero01,Pava09}. It is natural to know whether a fully open
eruption is possible in the multipolar dominated background. We
now investigate the energy accumulation process of the flux rope
in multipole dominated background with $a_1=2/3$, the field
configuration of which is also shown in Fig. \ref{M1d5}. The
detailed result is shown in the right panel of Fig. \ref{W0}. The
accumulated energy, $W_{\rm pre}(h_c)/W_{\rm pot}$, in the
critical pre-eruptive state is about $1.04$. However, energy
threshold constrained by the fully open state, $W^{\rm f}_{\rm
open}/W_{\rm pot}$, is $2.090$. It is clear that this flux rope
eruption is impossible to fully open up the multipole dominated field.

However, it is interesting to note that the eruption may just
involve partial opening up of the closed flux systems for
multipole dominated background. It is conceivable that the partial
opening up of the background field requires less work to be done
than the full opening up of background field. This will reduce the
energy threshold constrained by the full opening up eruption.
Along this line we could find an alternative approach of the flux
rope eruption. For this type of eruption, it may just involve
partial opening up of the magnetic field, which has a lower
threshold of $1.024$ (shown as dashed-dotted line in the right
panel). It reduces the energy threshold constrained by fully open
field. The critical pre-eruptive state contains energy about
$1.5\%$ above the partially open threshold. The energy
accumulation in the magnetosphere becomes sufficient to drive a
partially open eruption.

\subsection{Further Numerical Results}

\begin{figure}
\includegraphics[scale=0.9]{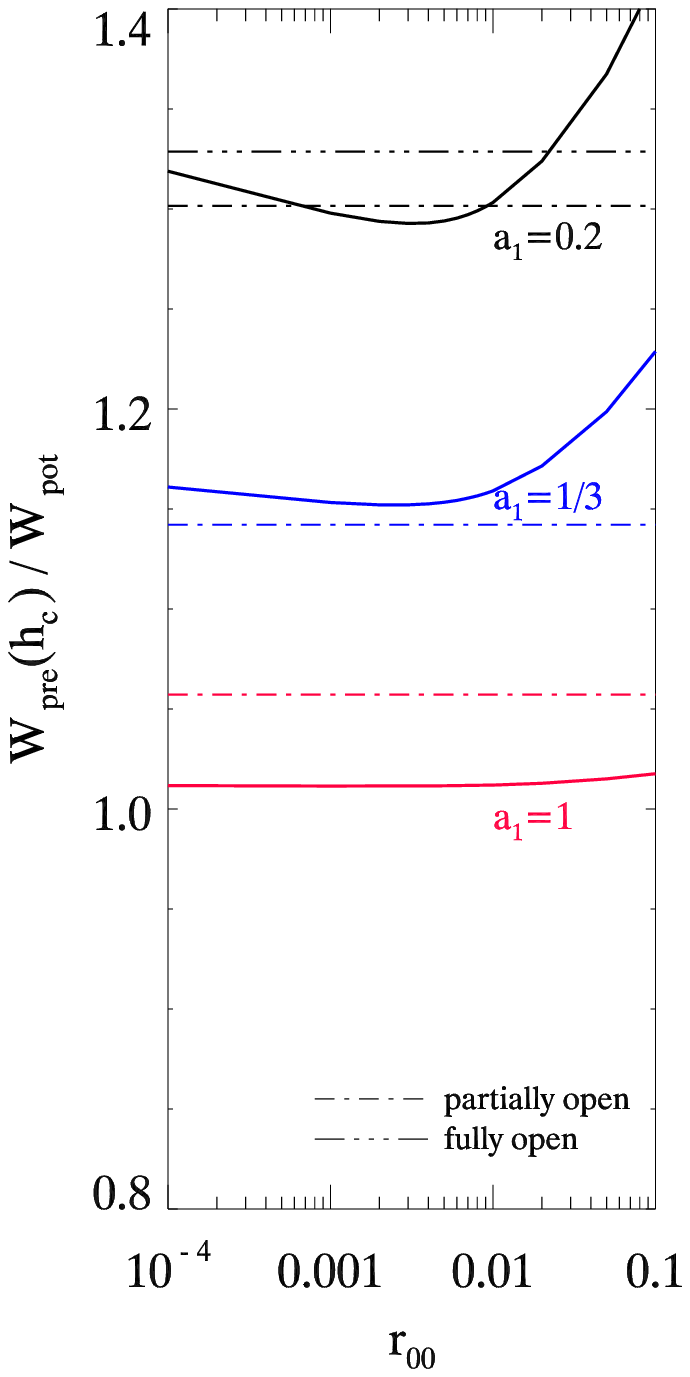}
\includegraphics[scale=0.9]{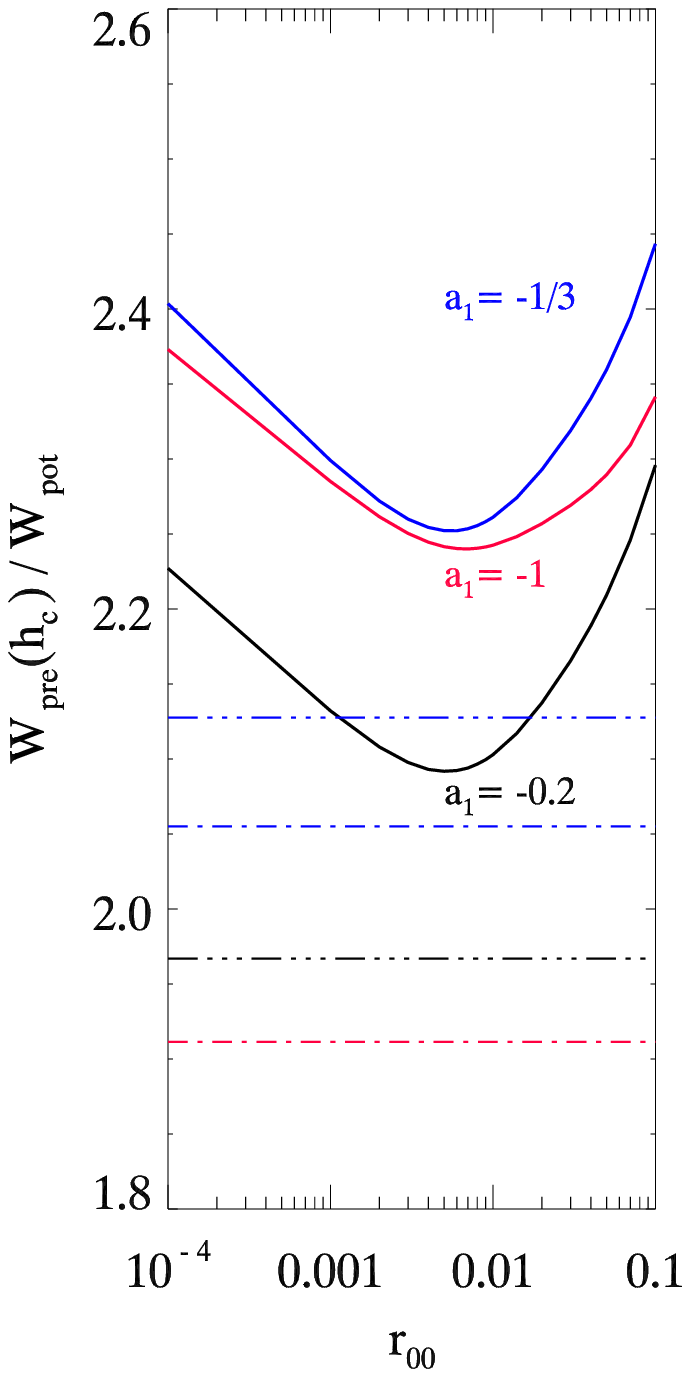}
\caption{\label{WWpot1} {\it Left}: Accumulated pre-eruptive
magnetic energy as functions of $r_{00}$ for centrally-caved field
with $a_1=0.2$ (black), $a_1=1/3$ (blue), and $a_1=1$ (red). The
corresponding threshold values of partially open fields are shown
in dashed-dotted lines in the same color. The corresponding
threshold values of fully open fields are shown in
dashed-three-dotted lines in the same color. {\it Right}: The same
as {\it left} panel but for centrally-arcaded field with
$a_1=-0.2$ (black), $-1/3$ (blue), and $-1$ (red).
 }
\end{figure}

Two representative examples are given in previous sections to show
the flux rope eruption from the perspective of energy budget. Here
we perform a comprehensive study on the flux rope energetics to
check the possibility of the flux rope's eruption in different
kind of background field. We investigate the accumulated magnetic energy,
$W_\mathrm{pre}(h_c)/W_\mathrm{pot}$, for different values of
$a_1\in[-1,1]$ and flux-rope radius $r_{00}\in[10^{-4},0.1]$. The
results are shown in Fig. \ref{WWpot1}.

Let's first focus on the cases with positive value of $a_1$, i.e.,
the cases with centrally-caved background field. In the left panel
of this figure, the black solid line represents the accumulated energy
$W_\mathrm{pre}(h_c)/W_\mathrm{pot}$ in a centrally-caved
multipolar field with $a_1=0.2$. Since the energy threshold of
both the fully and partially open fields are relevant in the
multipole dominated background, we show both the two energy
thresholds in the same panel for comparison. The threshold value
of the corresponding partially open field,
$W_\mathrm{open}^\mathrm{p}/W_\mathrm{pot}$, and the fully open
field, $W_\mathrm{open}^\mathrm{f}/W_\mathrm{pot}$, are shown
respectively in black dashed-dotted line and black
dashed-three-dotted line. Detailed calculation of the two energy
thresholds is given in Appendix A and B. It can be found that,
when $r_{00}\lesssim0.0007$ or $r_{00}\gtrsim0.01$, the
accumulated pre-eruptive energy could surpass the partially open
threshold. And when flux rope's minor radius is larger, i.e., with
$r_{00}\gtrsim0.02$, the accumulated pre-eruptive energy could
even surpass the fully open threshold. We also show the comparison
of the accumulated pre-eruptive energy (shown in blue solid line)
in a centrally-caved background field for the stronger multipolar
component $a_1=1/3$ with the relevant energy threshold.
The corresponding partially open threshold is shown in the blue
dashed-dotted line. It can be found that, for all possible value
of $r_{00}$, the accumulated pre-eruptive energy is higher than
the partially open threshold. However, none of them could surpass
the fully open threshold $\sim1.49$, which is beyond the scale of
this figure and not shown. We choose $a_1=1$ to represent a
centrally-caved background field with much stronger multipolar
component. The dependence of the accumulated pre-eruptive energy
on the flux rope minor radius for $a_1=1$ is shown as red solid
line. It is obvious that for all possible value of $r_{00}$, the
flux rope cannot accumulate enough energy to surpass the
corresponding partially open threshold, shown as the red
dashed-dotted line. Since the fully open threshold is always
higher than the partially open threshold, the flux rope could not
support fully open eruption either in this case. It is clear that
for centrally-caved multipole dominated field, if the multipole
components are too strong, the flux rope is not able to erupt.

Now let's turn to the centrally-arcaded background fields, i.e.,
the cases with the multipole parameter $a_1$ of negative sign. The
black solid line in the right panel of Fig \ref{WWpot1} represents
the accumulated pre-eruptive energy in a dipole dominated field
with $a_1=-0.2$. For the dipole dominated background, the relevant
energy threshold is only constrained by fully open field and this
threshold is shown in black dashed-three-dotted line in this
figure. It is obvious that, for all possible value of $r_{00}$,
the flux rope could build up more magnetic energy before the
catastrophe and give rise to fully open eruptions in the dipole
dominated background. The blue solid line in this figure
represents the accumulated pre-eruptive magnetic energy in a
centrally-arcaded multipolar field with $a_1=-1/3$. In this case,
the relevant thresholds are restricted by either the fully or the
partially open field. Detailed calculation shows that the
pre-eruptive energy could surpass both the corresponding partially
open threshold, shown in blue dashed-dotted line, and the fully
open threshold, shown in blue dashed-three-dotted line. The red
solid line represents the variation of the accumulated energy with
the flux rope's minor radius in a background field with $a_1=-1$,
i.e., a centrally-arcaded field with stronger multipolar
component. The pre-eruptive energy could surpass the corresponding
partially open threshold, shown in dashed-dotted red line.
However, the flux rope could not accumulate enough energy to get
over the corresponding fully open threshold ($\sim2.89$, not shown
in this figure).

\begin{figure}
\includegraphics[scale=0.9]{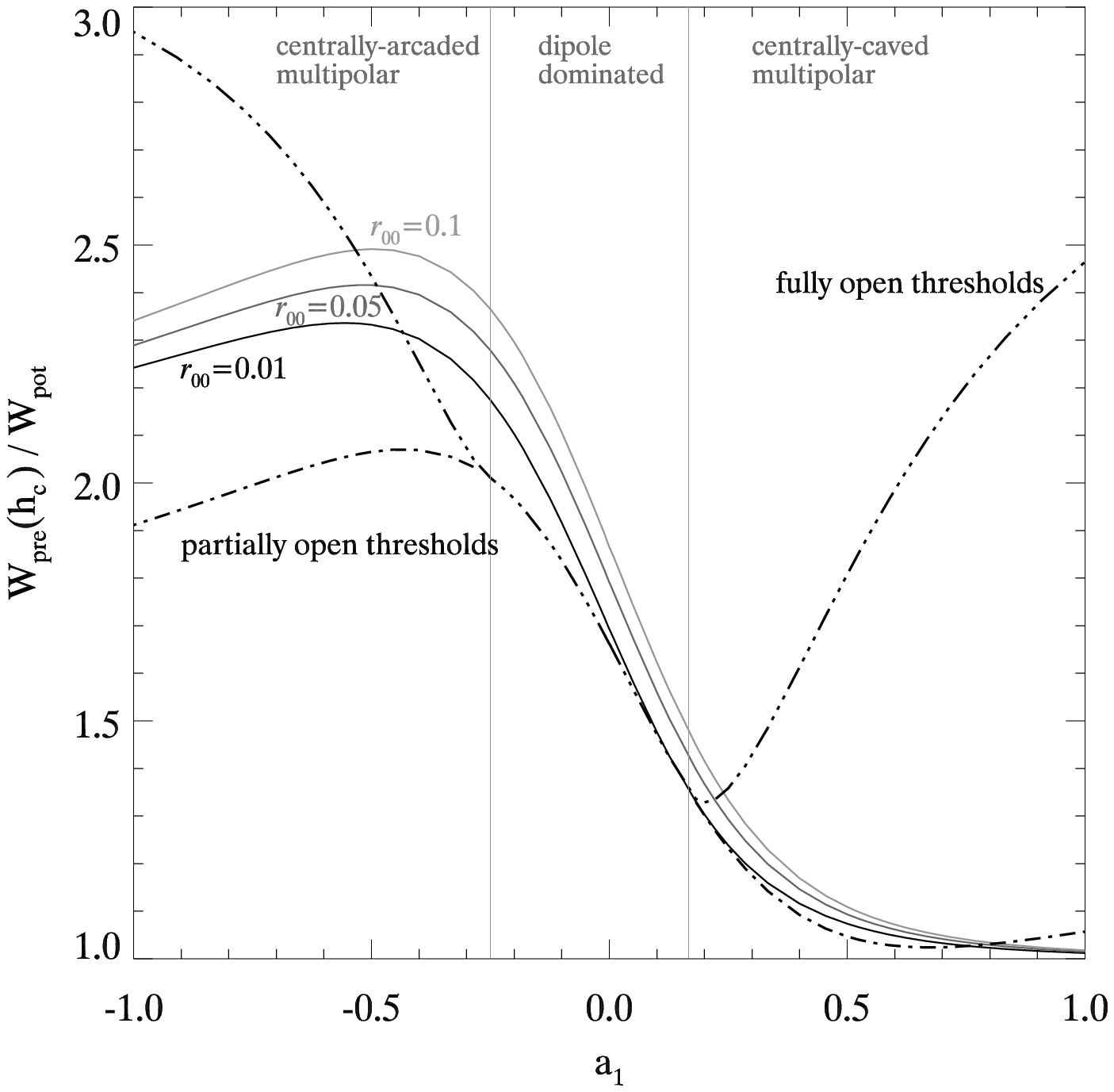}
\caption{\label{WWpot2} Accumulated pre-eruptive magnetic energy
as functions of $a_1$ with $r_{00}=0.01$ (black), $0.05$ (grey),
and $0.1$ (light grey). The corresponding threshold values of
partially-open fields and fully-open fields are shown in
dashed-dotted line and dashed-three-dotted line, respectively. Two
vertical lines represent the separatrix between the
dipole-dominated fields and the multipole-dominated fields.
 }
\end{figure}

Our prior results show that the possibility of the flux rope's
eruption depends on the parameter $a_1$ in a rather complex way.
These complex behavior can be more readily comprehensible in Fig.
\ref{WWpot2}. We show in this figure the dependence of accumulated
pre-eruptive magnetic energy on the parameter $a_1$ for different
fixed value of $r_{00}$. The relevant energy thresholds of fully
and partially open fields are shown as well in dashed-three-dotted
and dashed-dotted line, respectively. The comparison of the
accumulated pre-eruptive energy and the energy thresholds can be
made in a more straightforward way. The solid black, grey, and
light grey lines represent the accumulated pre-eruptive magnetic
energy for typical flux rope minor radius $r_{00}=0.01$, $0.05$, and $0.1$, respectively.
Note that partially open thresholds do not exit in dipole
dominated fields with $a_1\in[-1/4,1/6]$, which is shown as two
vertical lines in this figure.

We found that, in the dipole dominated background fields, i.e.,
$a_1\in[-1/4,1/6]$, it is always possible to drive a fully open
eruption. However, in most cases of the multipole background
fields, the fully open energy threshold is always greater than the
energy stored in the critical pre-eruptive state and it is
impossible for the flux rope to induce fully open eruptions.
Only in some special cases, when the multipole components are not so strong ,
i.e., $-0.55\lesssim a_1\lesssim-0.25$ or $1/6\lesssim
a_1\lesssim0.25$, it is possible to induce fully open eruptions.

More importantly, we note that in these fields, the eruption may
just involve partially opening up of the close the magnetic flux,
which provides an alternative approach for the flux rope
eruptions. It is clearly discernable in this figure that, the
energy release fraction, $\left[W_{\rm pre}(h_c)-W^{\rm p}_{\rm
open}\right]/W_{\rm pre}(h_c)$, is about $10\%\sim25\%$ during the
partially open eruptions in the centrally-arcaded backgrounds,
which is able to release and drive the giant flares. The energy
release fraction for flux ropes embedded in the centrally-caved
backgrounds is in a smaller range, within $\sim5\%$.
Note that there exits a special class of background fields with very strong
centrally-caved multipole component, $a_1\gtrsim0.75$. The
pre-eruptive energy possessed in flux ropes is always lower than
the partially open thresholds. These kind of background fields
cannot be opened up by eruptions of flux ropes.

Note in addition that the simplest case of $a_1=0$ has been
investigated by \citet{Lin98}. However, their approximate
treatment of the flux frozen constraints has led them to the
inappropriate result that the flux rope can not support the fully
open eruption. In our paper, we rigorously take into account the
flux-frozen constraint and come to the different results with
theirs\footnote{To double check our results, we also adopted the
approximation in Lin (1998) and reproduced their results.}.


\section{Conclusions and Discussions}

We propose a force-free magnetospheric model with an embedded
helically twisted flux rope. With the gradual variations at the
magnetar surface, the flux rope evolves quasi-statically in stable
equilibrium states. Upon the loss of equilibrium point is reached,
the global magnetospher is then destabilized and the flux rope
erupts catastrophically. During the process, the original closed
flux systems would be opened up, accompanied by rapid release of
the magnetic energy stored in magnetosphere. This energy release
is of vital importance for the outbursts of magnetars. However,
the feasibility that the flux systems' opening-up could be
achieved depends on whether the amount of energy accumulated prior
to the flux rope eruptions could surpass the energy thresholds
constrained by the post-eruptive open magnetic topologies.

In this paper, we adopt boundary conditions, which include both
the contribution from a dipolar component and a high order
multipolar component, to illustrate the complicated geometry of
magnetic field close to magnetar surface. We establish fully open
field for the dipole dominated magnetic fields, which involves the
opening up the single closed flux system in the backgrounds.
For multipole dominated closed background, we establish the
partially open field, which involves opening up of part of the
closed flux systems, as well as the fully open field, which
involves opening up of all closed flux systems. The opening up of
closed magnetic fields requires certain amount of work to be done
to overcome the attractive magnetic tension force. Since partially
open fields require less closed flux systems to be opened up, the
energy thresholds constrained by the partially open fields is
lower than those by the fully open fields. Both the field
configuration and the magnetic energy threshold of the two kinds
of open field are examined. Then we carefully investigate the
magnetic energy accumulation process before the catastrophe,
especially the magnetic energy stored at the critical catastrophic
point.

We find that it is possible to fully open up dipole dominated
background fields for catastrophic eruptions of flux ropes.
However, it is generally difficult to fully open up
multipole dominated background fields. In most cases with
multipole dominated backgrounds, the magnetic energy stored at
critical pre-eruptive point is significantly lower than the fully
open thresholds, which suggests that the flux rope can not support
fully open eruptions. Fortunately, we find that the accumulated
magnetic energy at the critical point is higher than the partially
open thresholds. This provides an alternative opportunity for the
flux rope to erupt in the multipolar magnetosphere.
The multipole dominated fields can be either centrally-caved or
centrally-arcaded, depending on the flux profiles on the magnetar
surface. Generally speaking, the magnetic energy stored in
critical pre-eruptive magnetosphere surpasses the partially open
energy threshold about $10\%\sim25\%$, if the flux rope is
initially embedded in a centrally-arcaded background field. For a
flux rope initially embedded in a centrally-caved background
field, the magnetic energy stored in critical pre-eruptive
magnetosphere could surpass the partially open threshold, if the
multipolar component is mildly strong, i.e.,
$1/6<a_1\lesssim0.75$. The energy release fraction is within
$\sim5\%$. If the multipolar component becomes even stronger,
$a_1\gtrsim0.75$, the accumulated magnetic energy cannot go beyond
the partially open threshold and the partially open eruption of
flux rope in not possible.

The magnetic energy of the critical pre-eruptive state in excess
of the fully or partially open threshold is assumed to be released
in fast dynamical timescale. Observationally, the total magnetic energy in the
magnetosphere is about $\sim 10^{46} ({\rm B}/10^{14}G)^2
(r_s/10{\rm km})^3$ergs, the giant flare is typically
$10^{44}$ergs, so only $1\%$ of magnetic energy release in the
magnetosphere could account for the giant flares. Theoretically,
all cases with surplus energy fraction larger than $1\%$ are
possible to drive magnetar giant flares. Specifically for boundary
conditions adopted in this paper, most cases with $-1\le
a_1\lesssim0.75$ are possible to drive magnetar giant flares.

In addition to the physical processes considered in this paper,
the magnetic field can also be opened up by the strong neutron
star wind (Bucciantini et al. 2006). If the field lines are opened
up by the neutron star wind at a larger distance from the neutron
star, the amount of open magnetic flux would be reduced, and the
magnetic energy to support the eruption would also decrease.
To study how the neutron star wind affects the flux rope eruption
energetics, we need to establish a model with a current sheet in
the magnetosphere. Currently, we are now trying to construct a
magnetosphere model to account for this important process. We
leave the investigation about the situations with current sheet
formation in a companion paper \citep{HYprep}.

\acknowledgments  We are grateful to the anonymous referee¡¯s
insightful comments, which improve this paper. 
This work has been supported by National Natural Science Foundation
of China (Grants 11203055, 10703012, 11173057, 11373064, 11121062 and 11173046), 
Open Research Program in Key Lab for the Structure and Evolution of Celestial Objects (Grant OP201301), 
Yunnan Natural Science Foundation (Grant 2012FB187), and Western Light Young Scholar Programme of CAS. 
This work is partly supported by the Strategic Priority Research Program "The Emergence of Cosmological Structures" of the Chinese Academy of Sciences (Grant No. XDB09000000) 
and the CAS/SAFEA International Partnership Program for Creative Research Teams. Part of the computation is performed at HPC Center, Yunnan Observatories, CAS, China.

\clearpage

\appendix

\section{Configurations of Partially Open and Fully Open Magnetic Fields}

\begin{figure}
\includegraphics[scale=0.8]{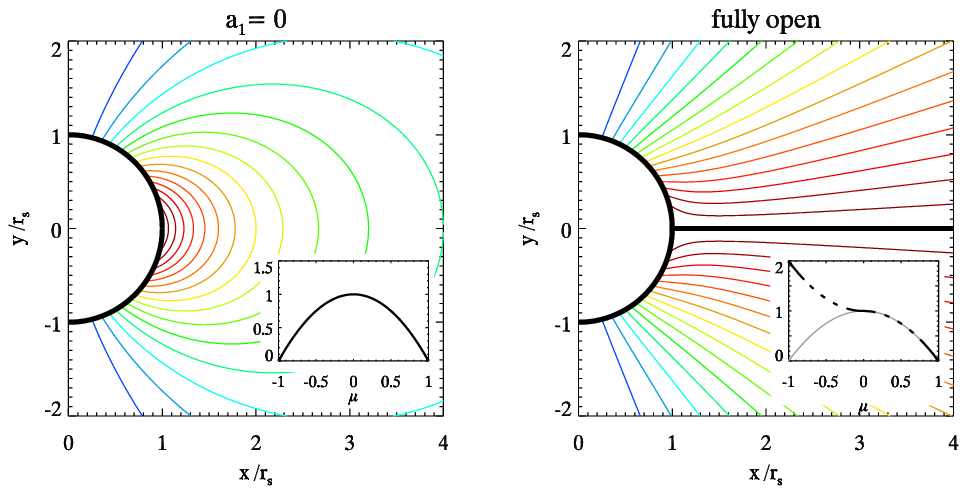}
\includegraphics[scale=0.8]{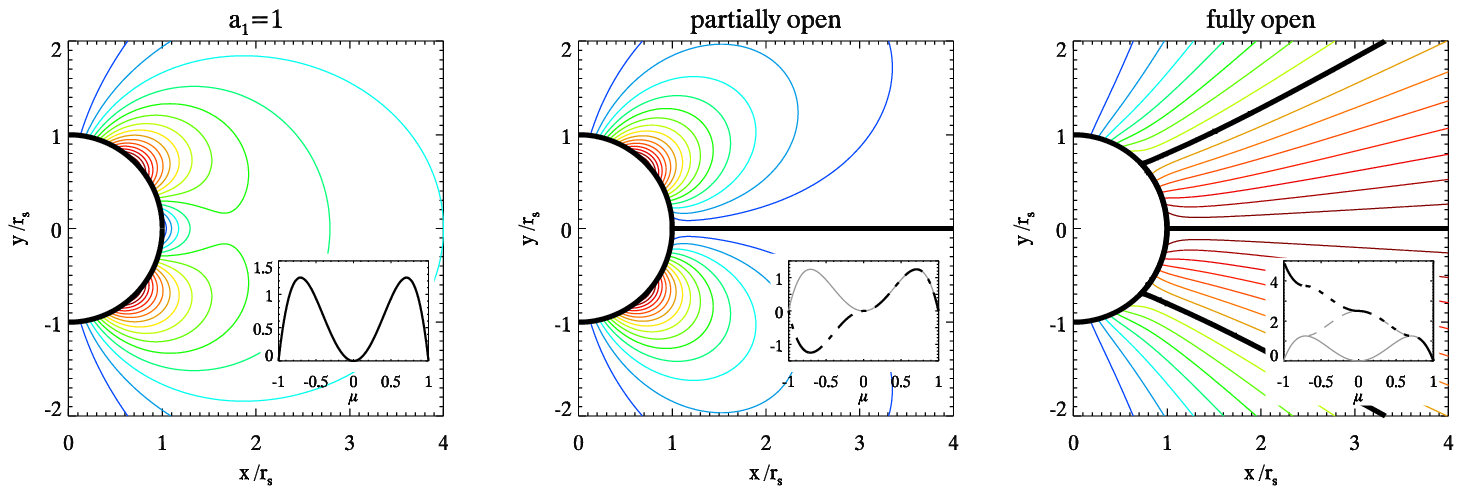}
\includegraphics[scale=0.8]{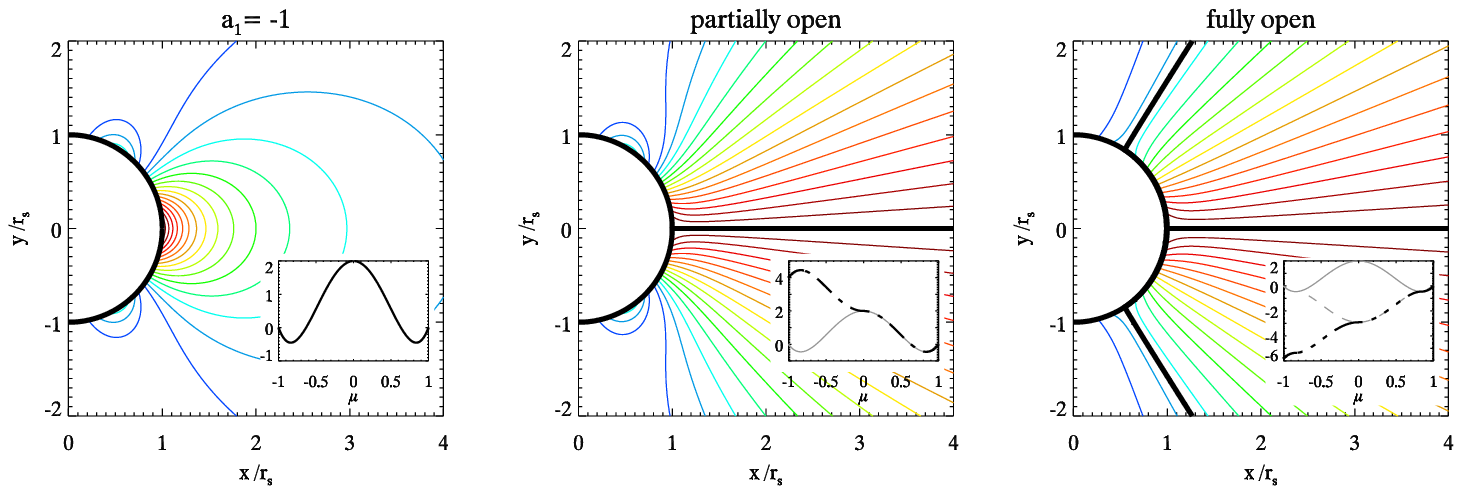}
\caption{\label{bound}{\it Top-Left}: Configuration of a dipolar
field. The boundary flux distribution is shown as solid line in
the sub-panel. {\it Top-Middle}: Configuration of the
corresponding fully open field. The thick solid line on the
equatorial plane represents the current sheet formed by the
opening up of the single closed flux system of the background. The
boundary flux distribution is obtained by flipping the surface
flux distribution of the potential field and is shown as
dashed-dotted line in the sub-panel. {\it Middle-Left}:
Configuration of a centrally-caved potential field with $a_1=1$.
{\it Middle-Middle}: Configuration of partially open field with $a_1=1$.
The thick solid line on the equatorial plane
represents the current sheet formed by the opening up of the central closed flux system.
{\it Midle-Right}: Configuration of fully open field with $a_1=1$,
with three current sheets formed by the opening up of all the
three closed flux systems in the background, shown in thick solid
lines. The boundary flux distribution is also obtained based on
the surface flux distribution of the potential field and is shown
as dashed-three-dotted line in the sub-panel. {\it Bottom-left}:
Configuration for a centrally-arcaded potential field with
$a_1=-1$. {\it Bottom-Middle} : partially open field with $a_1 =
-1$. {\it Bottom-Right}: fully open field with $a_1=-1$.
 }
\end{figure}

In this appendix, we describe procedures to get the fully open magnetic field
from dipole dominated field, and to get both fully and partially open field
from multipole dominated field. We show three illustrative examples in Fig. \ref{bound}.

In the upper row we show a dipolar field with $a_1=0$,
as an example of dipole dominated fields with $a_1\in[-1/4,1/6]$.
The configuration of potential background field is shown in top-left panel.
The thick solid semi-circle in these figures represents the magnetar surface.
To be clear, we show the boundary flux distribution in solid line in the sub-panel.
Only one extremum appears in boundary flux distribution.
The fully open field is obtained by simply flipping the surface flux distribution of
the potential field in the southern hemisphere, i.e., $-1\le\mu<0$ or $\pi/2<\theta\le\pi$.
The configuration of the fully open field is shown in top-middle panel.
The corresponding boundary flux distribution is shown in dashed-three-dotted line in the sub-panel.
The boundary flux distribution of the original closed field is also
shown in solid grey line for comparison.

There would appear three extremum points in the boundary flux distribution for parameters
$a_1\in[-1,-1/4)$ or $a_1\in(1/6,1]$, so that multipolar
configurations arise in the background magnetic field.
Under these circumstances, two kinds of open field configurations, i.e.,
partially open and fully open fields, can be obtained.
The middle row is for a centrally-caved field with $a_1=1$ and the lower row is for a
centrally-arcaded field with $a_1=-1$. Configurations of potential
background fields, partially open fields, and fully open fields are
shown in the left, middle and right panels, respectively.
In the following we describe the mathematical manipulations
to obtain these two kinds of open fields of the case $a_1=1$ as an example.
The case of $a_1=-1$ can be obtained in a similar way.

The partially open field is obtained by simply flipping the
surface flux distribution of the potential field in the southern
hemisphere. The corresponding boundary flux distribution is shown in dashed-dotted line
in the sub-panel of middle-middle panel.
It can be readily identified that only field lines near the
central extremum around $\mu=0$ are opened up, while the other two
closed flux systems around the nonzero extremum points remain
closed. In this sense, we call the resulting magnetic field
configurations as partially open fields.
The fully open field configurations can be obtained from original
potential field in two steps. The details are clearly illustrated
in the sub-panel of the middle-right panel. In the first step, the
boundary flux between the two nonzero extremum points are reversed.
After this step, the modified boundary flux contains only one
extremum point at $\mu=0$ (see the dashed line in the subpanel).
In the second step, the southern hemisphere boundary flux are
flipped and the resulting boundary flux distribution is shown as
the dashed-three-dots line. The closed configurations of the
patterns caused by the high order multipolar terms are also opened
up, shown in thick black lines near two poles of magnetar. There
are three current sheets in total in fully open field, together
with the equatorial one.

\newpage
\section{Energy of Partially Open and Fully Open Magnetic Fields}

The boundary condition of the post-eruptive partially open field
is obtained by flipping the flux function according to the
original boundary condition of the closed potential field
\citep{Yu11b}. Explicitly, the modified boundary flux distribution
of the partially open field can be written as (see the sub-panels
of partially open fields shown in Fig. 6)
\begin{eqnarray}
    \Theta^{\rm p}_{\rm open}&=&\left\{\begin{array}{cc}
        \Theta(\mu), & 0\le\mu\le1\\
        2\Theta(0)-\Theta(\mu), & -1\le\mu\le0 \ ,
        \end{array}\right.
\end{eqnarray}
where $\Theta(\mu)$ is already defined in Equation (\ref{pb}). The
general solutions to the GS equation are of the form
\begin{eqnarray}\label{a2k}
\Psi^\mathrm{p}_\mathrm{open}(r,\mu) &=& \Psi(r_s,0) (1-\mu) +
\sum_{k=1}^\infty a_{2k}r^{2k+1}\
\left[\frac{P_{2k-1}(\mu)-P_{2k+1}(\mu)}{4k+1}\right]\ .
\end{eqnarray}
To determine the partially open field, we have to specify the
coefficients $a_{2k}$'s in the above equation. For convenience, we
define the following flux function as
\begin{eqnarray}
    \Phi(r,\mu)&=&\Psi^\mathrm{p}_\mathrm{open}(r,\mu)\ -\ \Psi(r_s,0) (1-\mu)\nonumber\\
    &=& \sum_{k=1}^\infty a_{2k}r^{2k+1}\ \left[\frac{P_{2k-1}(\mu)-P_{2k+1}(\mu)}{4k+1}\right].
\end{eqnarray}
According to the orthogonality of associated Legendre polynomials $P_{2k}^1(\mu)$
and $\Psi(r_s,\mu)=\Psi_0\sigma\Theta(\mu)$, we can determine the coefficients $a_{2k}$ as
\begin{eqnarray}
    a_{2k}&=& \frac{4k+1}{r_s^{2k+1}}\ \int_0^1\Phi(r_s,\mu)\ P_{2k}^1(\mu)\ d\mu \nonumber\\
    &=&\Psi_0\sigma\ \frac{4k+1}{r_s^{2k+1}}\ \int_0^1\left[\Theta(\mu)-\Theta(0)(1-\mu)\right]\ P_{2k}^1(\mu)\ d\mu\ .
\end{eqnarray}
Once these coefficients are fixed, we can get the the partially
field configurations.

The boundary condition of the post-eruptive fully open field is
obtained in a similar way to the partially open field. The
difference is that we also need to flip the original boundary flux
profile in the range $[0,1]$. Hereafter we use $\Theta^*(\mu)$ to
denote the flipped boundary flux profile in the range $[0,1]$. The
modified boundary flux profile in the entire range $[-1,1]$ can be
expressed in terms of $\Theta^*(\mu)$ as (see the sub-panels of
fully open fields shown in Fig. 6)
\begin{eqnarray}
    \Theta^{\rm f}_{\rm open}&=&\left\{\begin{array}{cc}
        \Theta^*(\mu), & 0\le\mu\le1\\
        2\Theta^*(0)-\Theta^*(\mu), & -1\le\mu\le0 \ .
        \end{array}\right.
\end{eqnarray}
where the new surface flux distribution $\Theta^*(\mu)$ in the
range $[0, 1]$ is flipped as follows,
\begin{eqnarray}
\label{pbsh}
    \Theta^*(\mu) &=& \left\{\begin{array}{cc}
        \Theta(\mu)\ , & \mu_1\le|\mu|\le1 \\
        (4a_1+1)^2/(10a_1)\ -\ \Theta(\mu)\ , & 0\le|\mu|<\mu_1 \\
    \end{array}  \right.\ ,
\end{eqnarray}
where $\mu_1=\sqrt{(6a_1-1)/(10a_1)}$ is the nonzero extremum
point in the range $[0,1]$. Similarly, we can determine the
coefficients $a_{2k}$'s in Equation (\ref{a2k}) in terms of the
surface flux distribution of the fully open field, $\Theta^{\rm
f}_{\rm open}$. The fully open field subsequently can be
determined in complete detail.

The magnetic energy possessed in the post-eruptive state,
$W_\mathrm{open}^\mathrm{p}$ for partially-open fields state or
$W_\mathrm{open}^\mathrm{f}$ for fully-open fields state in this
paper, reads
\begin{eqnarray}
W^{\rm p,f}_\mathrm{open} &=& \int \frac{{{\bf B}_{\rm open}^{\rm
p,f}}^2}{8\pi}\ dV = \int _{\partial V} \frac{\left(B_{\rm
open}^{\rm p,f}\right)^2}{8\pi}({\bf r} \cdot d{\bf S}) -
\frac{1}{4\pi} \int_{\partial V} ({\bf B}_{\rm open}^{\rm
p,f}\cdot {\bf r}) ({\bf B}_{\rm open}^{\rm p,f}\cdot d{\bf S})\ ,
\end{eqnarray}
according to magnetic virial theorem.
The energy thresholds in fully-open fields and
partially-open fields are calculated by
$W_\mathrm{open}^\mathrm{f}/W_\mathrm{pot}$ and
$W_\mathrm{open}^\mathrm{p}/W_\mathrm{pot}$, respectively.

\newpage



\begin{thebibliography}{99}

    \bibitem[\protect\citeauthoryear{Aly}{1984}]{Aly84}
        Aly, J. J., 1984, ApJ, 283, 349

    \bibitem[\protect\citeauthoryear{Aly}{1991}]{Aly91}
        Aly, J. J., 1991, ApJ, 375, L61

    \bibitem[\protect\citeauthoryear{Antiochos et al.}{1999}]{Anti99}
        Antiochos, S. K., DeVore, C. R., \& Klimchuk, J. A., 1999, ApJ, 510, 485

    \bibitem[\protect\citeauthoryear{Beloborodov}{2009}]{Belo09}
        Beloborodov, A. M., 2009, ApJ, 703, 1044

    \bibitem[\protect\citeauthoryear{Bucciantini et al.}{2006}]{Bucc06}
        Bucciantini, N., Thompson, T. A., Arons, J., Quataert, E., \& Del Zanna, L., 2006, MNRAS, 368, 1717

    \bibitem[\protect\citeauthoryear{Chandrasekhar}{1961}]{Chandra61}
        Chandrasekhar, S, 1961, Hydrodynamic and Hydromagnetic Stability (Oxford: Oxford Univ. Press)

    \bibitem[\protect\citeauthoryear{Duncan \& Thompson}{1992}]{DT92}
        Duncan, R. C., \& Thompson, C., 1992, ApJ, 392, L9

    \bibitem[\protect\citeauthoryear{Feroci et al.}{2001}]{Fero01}
        Feroci, M., Hurley, K., Duncan, R. C., \& Thompson, C., 2001, ApJ, 549, 1021


    \bibitem[\protect\citeauthoryear{Forbes \& Priest}{1995}]{FP95}
        Forbes, T. G. \& Priest, E. R. 1995, ApJ, 446, 377

    \bibitem[\protect\citeauthoryear{Forbes}{2010}]{Forb10}
        Forbes, T. G., 2010, in Heliophysics: Space Storms and Radiation: Causes and
Effects, ed. J. S. Carolus \& L. S. George (Cambridge: Cambridge Univ.
Press), 159

    \bibitem[\protect\citeauthoryear{Gaensler et al.}{2005}]{Gaen05}
        Gaensler, B. M., Kouveliotou, C., \& Gelfand, J. D. 2005, Nature, 434, 1104

    \bibitem[\protect\citeauthoryear{Gavriil, Kaspi, \& Woods}{Gavriil et al.}{2002}]{GKW02}
        Gavriil, F. P., Kaspi, V. M., Woods, P. M. 2002, Nature, 419, 142

    \bibitem[\protect\citeauthoryear{Gill \& Heyl}{2010}]{GH10}
        Gill, R., \& Heyl, J. S. 2010, MNRAS, 407, 1926

    \bibitem[\protect\citeauthoryear{G\"{o}tz,  Mereghetti, \& Hurley}{G\"{o}tz et al.}{2007}]{GMH07}
        G\"{o}tz, D.,  Mereghetti, S., \& Hurley, K. 2007, Ap\&SS, 308, 51


     \bibitem[\protect\citeauthoryear{Huang \& Yu}{in prep}]{HYprep}
        Huang, L. \& Yu, C., in preparation

    \bibitem[\protect\citeauthoryear{Hurley et al.}{2005}]{Hurl05}
        Hurley, F., et al. 2005, Nature, 434, 1098

    \bibitem[\protect\citeauthoryear{Jones}{2003}]{Jone03}
        Jones, P. B. 2003, ApJ, 595, 342

    \bibitem[\protect\citeauthoryear{Klu\'{z}niak \& Ruderman}{1998}]{KR98}
        Klu\'{z}niak, W. \& Ruderman, M., 1998, ApJ, 505, L113

    \bibitem[\protect\citeauthoryear{Kouveliotou et al.}{1998}]{Kouv98}
        Kouveliotou, C., et al. 1998, Nature, 393, 235

    \bibitem[\protect\citeauthoryear{Levin \& Lyutikov}{2012}]{LL12}
        Levin, Y., \& Lyutikov, M. 2012, MNRAS, 427, 1574L

    \bibitem[\protect\citeauthoryear{Lin et al.}{1998}]{Lin98}
        Lin, J., Forbes, T. G., Isenberg, P. A., \& D\'{e}moulin, P. 1998, ApJ, 504, 1006

    \bibitem[\protect\citeauthoryear{Lundquist}{1950}]{Lund50}
        Lundquist, S. 1950, Ark. Fys., 2, 361


    \bibitem[\protect\citeauthoryear{Lyutikov}{2006}]{Lyut06}
        Lyutikov, M. 2006, MNRAS, 367, 1602

    \bibitem[\protect\citeauthoryear{Lyutikov}{2013}]{Lyut13}
        Lyutikov, M. 2013, arXiv: 1036.2264

    \bibitem[\protect\citeauthoryear{Mazets et al.}{1979}]{Maze79}
        Mazets, E. P., et al. 1979, Nature, 282, 587


    \bibitem[\protect\citeauthoryear{Mereghetti}{2008}]{Mere08}
        Mereghetti, S. 2008, A\&AR, 15, 225

    \bibitem[\protect\citeauthoryear{Mereghetti}{2013}]{Mere13}
        Mereghetti, S. 2013, in Proc. 26th Texas Symp. on Relativistic Astrophysics (arXiv: 1304.4825)

    \bibitem[\protect\citeauthoryear{Mereghetti \&  Stella}{1995}]{MS95}
        Mereghetti, S. \&  Stella, L. 1995, ApJ, 442, L17

    \bibitem[\protect\citeauthoryear{Palmer et al.}{2005}]{Palm05}
        Palmer, D. M., et al. 2005, Nature, 434, 1107

    \bibitem[\protect\citeauthoryear{Parfrey et al.}{2013}]{Parf13}
        Parfrey, K., Beloborodov, A. M., Hui, L., 2013, ApJ, 774, 92

    \bibitem[\protect\citeauthoryear{Pavan et al.}{2009}]{Pava09}
        Pavan, L., Turolla, R., Zane, S., \& Nobili, L. 2009, MNRAS, 395, 753

    \bibitem[\protect\citeauthoryear{Perna \& Gotthelf}{2008}]{PG08}
        Perna, R. \& Gotthelf, E. V. 2008, ApJ, 681, 522

    \bibitem[\protect\citeauthoryear{Priest \& Forbes}{2000}]{PF00}
        Priest E., \& Forbes T. 2000, Magnetic Reconnection. MHD Theory and Applications. Cambridge Univ. Press, Cambridge

    \bibitem[\protect\citeauthoryear{Ruderman}{1991}]{Rude91}
        Ruderman, M. 1991, ApJ, 366, 261

    \bibitem[\protect\citeauthoryear{Shafranov}{1966}]{Shaf66}
        Shafranov, V. D. 1966, Rev. Plasma Phys., 2, 103

    \bibitem[\protect\citeauthoryear{Sturrock}{1991}]{Stur91}
        Sturrock, P. A. 1991, ApJ, 380, 655



    \bibitem[\protect\citeauthoryear{Thompson, Lyutikov, \& Kulkarni}{Thompson et al.}{2002}]{TLK02}
        Thompson, C., Lyutikov, M., \& Kulkarni, S. R. 2002, ApJ, 574, 332


    \bibitem[\protect\citeauthoryear{Wood et al.}{2001}]{Wood01}
        Woods, P. M. et al. 2001, ApJ, 552, 748

    \bibitem[\protect\citeauthoryear{Woods \& Thompson}{2006}]{WT06}
        Woods, P. M., \& Thompson, C., 2006, in Compact Stellar X-Ray Sources, ed. W. H. G. Lewin \& van der Klis (Cambridge Univ. Press), 547

    \bibitem[\protect\citeauthoryear{Woods et al.}{2007}]{Wood07}
        Woods, P. M., Kouveliotou, C., Finger, M. H., et al. 2007, ApJ, 654, 470


    \bibitem[\protect\citeauthoryear{Yu}{2011}]{Yu11b}
        Yu, C., 2011, ApJ, 738, 75

    \bibitem[\protect\citeauthoryear{Yu}{2012}]{Yu12}
        Yu, C., 2012, ApJ, 757, 67

    \bibitem[\protect\citeauthoryear{Yu \& Huang}{2013}]{YH13}
        Yu, C. \& Huang, L., 2013, ApJ, 771, L46

    \bibitem[\protect\citeauthoryear{Zhang \& Low}{2001}]{ZL01}
        Zhang, M., \& Low, B. C., 2001, ApJ, 561, 406

\end{thebibliography}
\end{document}